\documentclass[12pt]{article}

\usepackage{amsmath,amssymb,latexsym,cite,xy,curves}
\xyoption{matrix}\xyoption{arrow}
\newtheorem{conj}{Conjecture}
\def\mcurve{\closecurve(
  -100,0,  -90,-35,  -70,-60, -40,-78, -20,-84,
  0,-86,  20,-84,  40,-78,  70,-60, 90,-35,
  100,0, 90,35,  70,60, 40,78, 20,84,
  0,86,  -20,84,  -40,78,  -70,60, -90,35)}
\def\arw#1#2#3#4{{
    \def\xscale{#3}
    \def\yscale{#3}
    \def\xscaley{-#4}
    \def\yscalex{#4}
    \put(#1,#2){\curve(0,0, 2,0.7, 3,2)}
    \put(#1,#2){\curve(0,0, 2,-0.7, 3,-2)}
}}

\DeclareMathAlphabet\Scr{U}{rsf}{m}{n}
\DeclareMathAlphabet\mathbi{U}{cmr}{bx}{it}

\def\O{\Scr{O}}

\def\P{{\mathbb P}}

\def\R{{\mathbb R}}
\def\Z{{\mathbb Z}}

\def\Hom{\operatorname{Hom}}
\def\hom{\operatorname{hom}}

\def\Ext{\operatorname{Ext}}

\def\SU{\operatorname{SU}}

\def\Cone{\operatorname{Cone}}

\def\ch{\operatorname{\mathit{ch}}}
\def\td{\operatorname{\mathit{td}}}

\def\p{\partial}

\def\CY{Calabi--Yau}

\def\cF{{\Scr F}}
\def\cX{{\Scr X}}
\def\cY{{\Scr Y}}

\def\DC{\mathbf{D}}

\def\O{\mathcal{O}}

\begin{document}

\title{D-brane stability, geometric engineering, and monodromy in
the Derived Category}

\author{Robert L.~Karp\footnote{Email: \texttt{karp@physics.rutgers.edu}}\\[4mm]
\normalsize  Department of Physics, Rutgers University\\
\normalsize  136 Frelinghuysen Rd., Piscataway, NJ 08854 USA\\[8mm]
}
\date{}

\maketitle

\begin{abstract}
We discuss aspects of topological B-type D-branes in the framework of the  derived category of coherent sheaves $\DC(X)$ on a \CY\ 3-fold $X$. We analyze the link between massless D-branes and monodromies in the CFT moduli space. A classification of all massless D-branes at any point in the moduli space is conjectured, together with an associated monodromy. 
We test the conjectures in two independent ways. First we establish a composition formula for certain Fourier-Mukai functors, which is a consequence of the triangulated structure of  $\DC(X)$. 
Secondly, using $\pi$-stability we rederive the stable soliton spectrum of the pure ${\cal N}=2$ supersymmetric $\SU(2)$ Seiberg-Witten theory. In this approach, the simplicity of the spectrum rests on Grothendieck's theorem concerning vector bundles over $\P^1$.
\end{abstract}
\newpage


\section{Background}

The original idea that a D-brane is a subspace in target space where open strings can {\em end} is too simplistic. In the case of closed strings we know that even at $g_s=0$ there are $\alpha'$-corrections which modify the classical geometry. Accordingly, one expects that the picture of D-branes as vector bundles over submanifolds is to be altered as well.

On a \CY\ n-fold $X$, there are two types of boundary conditions \cite{Becker:1995kb,Ooguri:1996ck}, and hence D-branes: A-type and B-type. To gain additional  control one can consider a suitable topologically twisted version of the boundary non-linear sigma model. The topological twisting can itself be of A-type or B-type. Moreover, A-type branes are compatible with the A-twist, while  B-type branes are compatible with the B-twist. For topological D-branes the following picture emerges:
\begin{itemize}
\item A-type  -- wrapping Lagrangian real n-cycles, and giving rise to objects in the derived Fukaya category of Lagrangian submanifolds;
\item B-type  -- wrapping holomorphic cycles, and giving rise to objects in the bounded derived category of coherent sheaves $\DC(X)$ of $X$.
\end{itemize}
The homological mirror symmetry conjecture of Kontsevich gives a categorical equivalence between these two triangulated $A_\infty$-categories. The conjecture has already been proved for elliptic curves, and recently for quartic K3-surfaces \cite{Seidel:quartic}.

In this paper we focus on {\bf B-branes}, where using mirror symmetry, all the $\alpha'$-corrections can be understood exactly \cite{Aspinwall:2001dz}.
But now, instead of physical B-branes, living in the boundary conformal field theory, we have {\em topological B-branes}, and a priori it is not clear what is relationship between them. This is where the notion of {\em  $\Pi$-stability} pioneered by Douglas et al. \cite{Douglas:2000ah,Douglas:2000gi} comes into play: a topological B-brane is physical if it is $\Pi$-stable. We will have more to say  about $\Pi$-stability, but first we have to review some terminology. For the precise definition of $\Pi$-stability the reader should consult for example \cite{Aspinwall:2001dz}.

Given two D-branes, $\mathsf{A}$ and $\mathsf{B}$, the totality of topological
open strings stretching between them is represented by
$\Hom(\mathsf{A}[n],\mathsf{B})$, for any  integer $n$. Note that  for a D-brane $\mathsf{A}$, $\mathsf{A}[1]$ represents its anti-brane.

The most interesting open strings are in $\Hom(\mathsf{A}[-1],\mathsf{B})$.
Giving  {\bf vev} to such a string forms a potential bound state, depending on whether the   open string is tachyonic or not. Mathematically this gives us a cone, $\Cone(f:\mathsf{A}[-1]\to\mathsf{B})$, and  a distinguished triangle:
\begin{equation}\label{eq:phidef1}
\mathsf{A}[-1]\to\mathsf{B}\to\Cone(f)\to\mathsf{A}\,.
\end{equation}
This construction brings democracy to bound states: any vertex of this triangle is a potential bound state of the other two vertices.

Each stable D-brane is given a { grade} $\varphi\in\R$ which varies
continuously over the moduli space, although not in a single-valued way \cite{Douglas:2000ah,Douglas:2000gi}:
$  \varphi(\mathsf{A}) = -\arg(Z(\mathsf{A}))/\pi$ .
Near the large radius limit the central charge~is
\begin{equation}
  Z(\cF)=\int_X
  {\rm e}^{B+iJ}\ch(\cF)\sqrt{\td(X)}+\ldots \rm(quantum\, corrections).  \label{eq:LRL}
\end{equation}
Using { mirror symmetry} we can in fact evaluate all the quantum corrections \cite{Aspinwall:2001dz}. Then the mass of an open string becomes exactly computable as well: $M^2_{\mathsf{A}[-1]\to \mathsf{B}}\sim\varphi(\mathsf{B})-\varphi(\mathsf{A})$. It is $M^2_{\mathsf{A}[-1]\to \mathsf{B}}$ that determines the stability of $\Cone(f)$ in (\ref{eq:phidef1}), whether it is a bound state or not.

\section{Monodromy}

To learn more about the behavior of D-branes in non-trivial backgrounds, we will subject them to monodromy transformations. Henceforth we follow the set of stable D-branes as we traverse a loop in the moduli space of complexified Kahler forms. The interesting loops are those that cross a {line of marginal stability} for some D-brane $\mathsf{B}$, causing it to decay, and those that encircle points in moduli space where some D-brane $\mathsf{A}$ becomes massless. To restore the physics we had before looping we are required to {relabel} the D-branes. This relabeling is the monodromy action on D-branes.
Note that monodromy is not a statement about a certain D-brane manifestly  becoming
another D-brane, but rather it is the relabeling process.

A good place to start looping is the
{discriminant } locus $\Delta$.  This is the subvariety in the moduli space of complexified Kahler forms  where the  associated conformal field theory is singular. Alternatively, it is the locus in the moduli space of complex structures of the mirror \CY\ where this latter one becomes singular.

The paradigmatic example of monodromy is given by the following
\begin{conj}({Kontsevich, Horja, Morrison})
The monodromy action for looping around a component of the discriminant locus where a {single } D-brane $\mathsf{A}$ becomes
massless  is: $\mathsf{B}\longmapsto\Cone(\hom(\mathsf{A},\mathsf{B})\otimes\mathsf{A}
\to\mathsf{B}).$
\end{conj}
This monodromy can be understood physically, at least for some branes $\mathsf{B}$, as $\mathsf{B}$ splitting off $\mathsf{A}[1]$'s a certain number of times \cite{en:Horja}.
Naturally the question arises: what could $\mathsf{A}$ be? An instance is given by
\begin{conj}
At a generic point on the {primary component} of the discriminant locus $\Delta$, it is  the {D6-brane
$\O_X$} wrapping the 3-fold, and its translates, that become
massless. At a generic point no other D-branes become massless.
\end{conj}
But can we do better, deal with more {interesting} degenerations? Maybe several D-branes becoming massless? Possibly infinitely many becoming massless?
The typical situation as we approach a {wall} of
the K\"ahler cone is that some subspace $E$ of the \CY\ $X$ shrinks to a subspace $Z$:
\begin{equation}\label{ez}
\xymatrix{
E\, \ar[d]^q\ar@{^{(}->}[r]^i &X\\
Z
}
\end{equation}
Following \cite{Horj:EZ} we call this an EZ-degeneration. In this case the following conjecture is known:
\begin{conj} \cite{en:Horja}
Any D-Brane which becomes massless at a point on the
discriminant associated with an EZ-degeneration is generated by objects
of the form $i_*q^*\mathsf{z}$ for $\mathsf{z}\in\DC(Z)$.
  \label{c:massless}
\end{conj}

In such a case,  there is usually a rational
curve $\P^1$ in the moduli space, connecting the two limit points: the large radius limit point with the EZ-point. The EZ-point corresponds to the phase reached by following the EZ-degeneration through the {wall} of the Kahler cone. Regarding this situation we have the following
\begin{conj}\cite{en:Horja,Horj:EZ}\label{c:4}
The monodromy around the discriminant inside the above $\P^1$ is the following
autoequivalence of the derived category of coherent sheaves on $X$, i.e. of $\DC(X)$:
 $ \mathsf{B} \mapsto \Cone(i_*q^*q_*i^!\mathsf{B}\to\mathsf{B})$.
\end{conj}

In what follows, we present two different ways of checking the consistency of these conjectures. The first example is based upon the fact that a certain monodromy can be computed in two different ways. Firstly by making use of the Conjectures, and secondly by exploring the simple structure of loops on marked  $\P^1$'s. The second example, based on geometric engineering, will also probe aspects of $\Pi$-stability. Nothing we say here is new, most of what follows can be found in \cite{en:Horja} and \cite{en:Paul}. The reader is also referred to these papers for most of the details.

\section{ Consistency checks}

\subsection{The Model}  \label{s:mod}

As a starting point we consider $X$ to be the well known degree 8 hypersurface in the resolution of {$\P^4_{\{2,2,2,1,1\}}$} (see e.g. \cite{Candelas:1994dm,Candelas:1994hw}). $X$ is then a {K3-fibration} $\pi\!:X\rightarrow C$ over a base $C\cong\P^1$, with a
section. Call the generic K3 fiber $F$.

What we want is the quantum-corrected moduli space of complexified
{ K\"ahler forms} on $X$. This can be understood via mirror symmetry as the {complex} structure moduli space of the mirror $Y$. Choosing {algebraic coordinates} on the moduli space, the ``primary'' component of the discriminant is given by
$\Delta_0 = (1-2^8x)^2 - 2^{18}x^2y.$ The K\"ahler cone is two dimensional, where $x$ controls the size of the K3-fiber, and $y$ the size of the base $C\cong\P^1$.

\subsection{ Consistency of different monodromies}

We can make the base $\P^1$ very large by sending $y\to0$. But the primary component $\Delta_0$ has
second order intersection with $y=0$, call this point $P$. The  $y=0$ curve is rational,
and has three distinguished points: the large \CY\ limit point ($x=y=0$), the
hybrid ``$\P^1$-phase'' point, and $P$.

We want to analyze the monodromy around $P$. The transition associated to $P$ is of EZ-type, and consists of collapsing $X$ onto the $\P^1$ base: $E=X$, $Z=\P^1$, $i=id_X$ and $q=\pi$. Using Conjecture \ref{c:4} we can immediately deduce the  monodromy action: $\mathsf{B}\mapsto\mathbi{H}(\mathsf{B}) = \Cone(\pi^*\pi_*\mathsf{B}\to\mathsf{B})$.

On the other hand singularity theory allows us to write the monodromy around $P$ in a completely different way (see e.g. \cite{navigation}), as
$\mathbi{L}^{-1}\mathbi{K}\mathbi{L}\mathbi{K}$, where $\mathbi{L}$ is the  large radius monodromy  $\mathbi{L}(\mathsf{B}) = \mathsf{B}\otimes\O_X(F)$, while $\mathbi{K}$ is the Kontsevich et al. monodromy of Conjecture 1: $  \mathbi{K}(\mathsf{B}) = \Cone(\hom(\O_X,\mathsf{B})\otimes\O_X     \to\mathsf{B})$.

By making extensive use of the structures in the derived categories of  $X$  and $\P^1$, it was shown in \cite{en:Horja} that $\mathbi{H}=\mathbi{L}^{-1}\mathbi{K}\mathbi{L}\mathbi{K}$ indeed holds, providing a strong consistency check for the monodromy conjectures.

\subsection{Geometric engineering}

Now we turn to the second test of our conjectures.
We can use the techniques of \cite{Kachru:1996fv,Katz:1997fh} to engineer an $N=2$ pure $\SU(2)$ gauge theory using the above mentioned \CY\ $X$ \cite{en:Paul}. The goal is to understand the stable BPS solitons spectrum of the {Yang--Mills}  theory \cite{SW:I,Bilal:1} from the $\Pi$-stable BPS D-branes spectrum of the {Type IIA} theory.

The geometric engineering limit requires again the base $C$  to be {very} large, $y\to0$, and we must also {lie} very close to the discriminant locus. Note that the second order intersection point $(2^{-8},0)$   splits for $y=\epsilon$. Introducing the variable $u$ by 
$$u^{-2}={4y}/(2^8 x-1 ) ^{2}$$ 
places the two intersection points at $u=\pm 1$. In this limit the Picard--Fuchs equations satisfied by the periods on $X$ reduce to the one satisfied by the Seiberg-Witten periods\cite{en:Paul}: 
$$(1-u^2)\p_u^2\,a-2u\p_u\,a-\frac{1}{4}\,a=0.$$ 
But it is precisely the periods that together with the D-brane charges determine the central charges, and hence the stability of solitons.

The  D-brane charge  is measured by K-theory:
$K^*(\P^1)=H^0(C)\oplus H^2(C)=\Z^2$. For convenience take $\O_C$ and $\O_p$, which also
generate $\DC(C)$, as the generators of $K^*(\P^1)$. Therefore, we require only two basic central charges:
$  a = Z(\pi^*\O_p),\,  a_D = Z(\pi^*\O_C)$.

Conjecture 3 and and the same  knot theoretic computation as in the previous subsection gives  the following assignments:
\begin{itemize}
\item {magnetic monopole}: {$\O_X=\pi^*\O_C$}, i.e., the 6-brane wrapping $X$;
\item {dyon}: {$\O_X(F)=\pi^*\O_C(1)$}, i.e., a 6-brane with 4-brane charge;
\item {W-boson}: {$\pi^*\O_p$}, i.e., a 4-brane wrapped around a K3 fiber.
\end{itemize}
The charge assignment is in agreement with the short exact sequence of sheaves
$$ 0\to\O_C(-1)\to\O_C\to\O_p \to 0,$$ 
and it's pullback to $X$. This is in itself a non-trivial fact. We can also read off the following identifications:
 magnetic charge = {\em rank\/},
 electric charge = {\em degree\/}.

\subsubsection{Weak coupling region}

For very large $|u|$ we are at the  large radius limit of $X$, where all the
$\alpha'$-corrections are very small. In this limit B-type D-branes
correspond to stable holomorphic vector bundles over subvarieties of $X$. Conjecture \ref{c:massless} allows us to work on $C\cong\P^1$ \cite{en:Paul}. For brevity we consider {only} the stability of {rank 2} bundles.

By Grothendieck's theorem, a rank $r$ vector
bundle $V$ over $C\cong\P^1$ is a  direct sum of line bundles:
$$V\cong \O_C(s_1) \oplus \O_C(s_2) \oplus\ldots\oplus \O_C(s_r).$$ 
Thus the general rank 2 bundles is of the form $\O_C(s)\oplus\O_C(t)$. But
$\O_C(s)\oplus\O_C(t)$ fits into the triangle
$$\O_C(s)\to\O_C(s)\oplus\O_C(t)\to\O_C(t)\to\O_C(s)[1],$$
where $f\in\Hom(\O_C(s),\O_C(t)[1])=\Ext^1(\O_C(s),\O_C(t))$ determines the extension class.

There are two cases to consider, depending on whether $f=0$ or not, resulting in the following \cite{en:Paul}: {\em the rank two bundle
$\O(s)\oplus\O(t)$ will decay into $\O(s')\oplus\O(t')$ for some $s',t'$ such that
$s'+t'=s+t$ and $|s'-t'|\leq1$.}
Note that for $\O(s)\oplus\O(t)$ with
$|s-t|\leq1$, $\Ext^1(\O_C(s),\O_C(t))=\Ext^1(\O_C(t),\O_C(s))=0$, so
there are no open strings to either stabilize or destabilize, and the
two constituents are free to drift apart.

Similarly, one proves that any dyon $(r,m)$ for $r>1$ is unstable to decay
into dyons with $r=1$, and there are no stable
states of charge $(0,m)$ for $m>1$ either \cite{en:Paul}, in line with \cite{SW:I,Bilal:1}.

\subsubsection{Strong coupling region}

By considering the distinguished triangle
$\O_C(m)\to\O_C(m+1)\to\O_p\to\O_C(m)[1]$ we can show that the only stable solitons are: {$\O_C$} and {$\O_C(1)$}, i.e., the monopole and the dyon, again in line with \cite{SW:I,Bilal:1}. In particular, the $W$-boson $\O_p$ decays into $\O_C(1)$ and $\O_C[1]$ \cite{en:Paul}.

\subsubsection{Returning to weak coupling}

Suppose now that we continue our journey, back to the weakly coupled region.
We would expect to recover the original set of stable solitons. But they may not be the same
elements of the derived category, as they might undergo monodromy.

One would expect that as we cross the MS-line, $\O_p$ would become stable again.
Unfortunately $M^2_{\O_C\to\O_C(1)}\sim \varphi(\pi^*\O_C(1))-\varphi(\pi^*\O_C[-1])$
gets larger and larger, so $\O_p$ never stabilizes.
Fortunately, to rescue us there come some new bound states! Define $\cX$ by the triangle
$$\pi^*\O_C(1)[-1]\to\pi^*\O_C[2]\to\cX\to\pi^*\O_C(1),$$ 
then iterate it:
$$\cY_m\to\cY_{m+1}\to\cX\to\cY_m[1].$$ 
It turns out that {\em all\/} of the $\cY_m$'s become stable as we pass  the MS-line back into the weak-coupling regime.

This infinite tower of new states $\cY_m$ is the ``replacement" for the old states $\O_C(m)$, of charge $(1,m)$. It is also true that under monodromy around $u=1$ (we can use Conjecture 1 for this) the set of states $\O_C(m)$ turn into the set of states $\cY_m$.

But there is still an {aesthetic discrepancy}: the
original set of weak coupling states have very clear geometric interpretation as wrapped branes, while after monodromy we have rather exotic objects.
However, from the derived category point of view they look quite
symmetric (see the sketch below): in the strong-coupling region we have only
two stable D-branes: $\mathsf{A}$ and $\mathsf{B}$. As
we move across the MS-line into the weak coupling regime they can form a bound state
 $\mathsf{W}$ which plays the r\^ole of the W-boson. Crossing  \textit{upwards} this is done by
$\mathsf{A}\to\mathsf{B}$, while crossing \textit{downwards} by
$\mathsf{B}\to\mathsf{A}[3]$. This is essentially Serre duality at
work. Observe also the {\bf 3} appearing, as if the solitons in Seiberg--Witten theory knew
that they should be associated with a  {3-fold}!
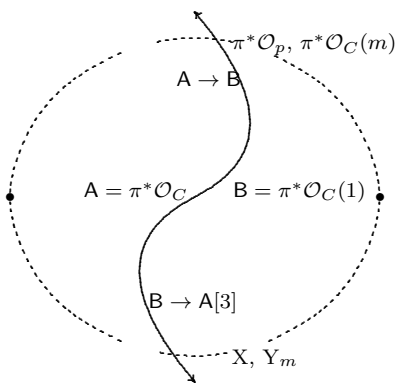
\setlength{\unitlength}{0.7pt}
\vspace{.51in}
\begin{center}
\begin{picture}(240,200)(-130,-100)
  \curvedashes{3,1}
  \mcurve
  \curvedashes{}
  \put(-100,0){\circle*{5}}
  \put(100,0){\circle*{5}}
  {\put(0,0){\curve(0,0, -30,-40, 0,-100)}}
    \arw{0}{-100}{-0.707}{0.707}
   {\put(0,0){\curve(0,0, 30,40, 0,100)}}
  \arw{0}{100}{0.707}{-0.707}
  \put(-10,60){\text{\scriptsize $\mathsf{A}\to\mathsf{B}$}}
  \put(20,80){\text{\scriptsize $\pi^*\O_p$, $\pi^*\O_C(m)$}}
  \put(-60,0){\text{\scriptsize
$\mathsf{A}=\pi^*\O_C\qquad\mathsf{B}=\pi^*\O_C(1)$}}
  \put(-25,-60){\text{\scriptsize $\mathsf{B}\to\mathsf{A}[3]$}}
  \put(20,-90){\text{\scriptsize $\cX$, $\cY_m$}}
\end{picture}
  \label{f:paths2}
\end{center}

\begin{flushleft}
\vspace{1in}

{\bf \large Acknowledgments}

I would like to thank my advisor Paul Aspinwall for his guidance, and  Paul Horja for enjoyable collaboration.  This work was supported in part by the NSF grant DMS-0074072.
\end{flushleft}


\newpage

\begin{thebibliography}{99}

\bibitem{Becker:1995kb}
K.~Becker, M.~Becker, and A.~Strominger,
\newblock {\em Five-branes, membranes and nonperturbative string theory},
\newblock Nucl. Phys. {\bf B456} (1995) 130--152, hep-th/9507158.

\bibitem{Ooguri:1996ck}
H.~Ooguri, Y.~Oz, and Z.~Yin,
\newblock {\em D-branes on Calabi-Yau spaces and their mirrors},
\newblock Nucl. Phys. {\bf B477} (1996) 407--430, hep-th/9606112.

\bibitem{Seidel:quartic}
P.~Seidel,
\newblock {\em Homological mirror symmetry for the quartic surface},
\newblock math.SG/0310414.

\bibitem{Aspinwall:2001dz}
P.~S. Aspinwall and M.~R. Douglas,
\newblock {\em D-brane stability and monodromy},
\newblock JHEP {\bf 05} (2002) 031, hep-th/0110071.

\bibitem{Douglas:2000ah}
M.~R. Douglas, B.~Fiol, and C.~Romelsberger,
\newblock {\em Stability and BPS branes},
\newblock hep-th/0002037.

\bibitem{Douglas:2000gi}
M.~R. Douglas,
\newblock {\em D-branes, categories and N = 1 supersymmetry},
\newblock J. Math. Phys. {\bf 42} (2001) 2818--2843, hep-th/0011017.

\bibitem{en:Horja}
P.~S. Aspinwall, R.~L. Karp, and R.~P. Horja,
\newblock {\em Massless D-branes on Calabi-Yau threefolds and monodromy},
\newblock Commun. Math. Phys. {\bf 259} (2005) 45--69, hep-th/0209161.

\bibitem{Horj:EZ}
R.~P. Horja,
\newblock {\em Derived category automorphisms from mirror symmetry},
\newblock Duke Math. J. {\bf 127} (No. 1)  (2005) 1--34,
  math\-.\-AG\-/\-0103231.

\bibitem{en:Paul}
P.~S. Aspinwall and R.~L. Karp,
\newblock {\em Solitons in Seiberg-Witten theory and D-branes in the derived
  category},
\newblock JHEP {\bf 04} (2003) 049, hep-th/0211121.

\bibitem{Candelas:1994dm}
P.~Candelas et~al.,
\newblock {\em Mirror symmetry for two parameter models. I},
\newblock Nucl. Phys. {\bf B416} (1994) 481--538, hep-th/9308083.

\bibitem{Candelas:1994hw}
P.~Candelas, A.~Font, S.~Katz, and D.~R. Morrison,
\newblock {\em Mirror symmetry for two parameter models. 2},
\newblock Nucl. Phys. {\bf B429} (1994) 626--674, hep-th/9403187.

\bibitem{navigation}
P.~S. Aspinwall,
\newblock {\em Some navigation rules for D-brane monodromy},
\newblock J. Math. Phys. {\bf 42} (2001) 5534--5552, hep-th/0102198.

\bibitem{Kachru:1996fv}
S.~Kachru et~al.,
\newblock {\em Nonperturbative results on the point particle limit of N=2
  heterotic string compactifications},
\newblock Nucl. Phys. {\bf B459} (1996) 537--558, hep-th/9508155.

\bibitem{Katz:1997fh}
S.~Katz, A.~Klemm, and C.~Vafa,
\newblock {\em Geometric engineering of quantum field theories},
\newblock Nucl. Phys. {\bf B497} (1997) 173--195, hep-th/9609239.

\bibitem{SW:I}
N.~Seiberg and E.~Witten,
\newblock {\em Electric-magnetic duality, monopole condensation, and
  confinement in N=2 supersymmetric Yang-Mills theory},
\newblock Nucl. Phys. {\bf B426} (1994) 19--52, hep-th/9407087.

\bibitem{Bilal:1}
F.~Ferrari and A.~Bilal,
\newblock {\em The Strong-Coupling Spectrum of the Seiberg-Witten Theory},
\newblock Nucl. Phys. {\bf B469} (1996) 387--402, hep-th/9602082.

\end{thebibliography}
\end{document}